\documentclass[preprint,journal]{vgtc}            

\onlineid{0}

\preprinttext{To appear in IEEE Transactions on Visualization and Computer Graphics (VIS 2024).}

\vgtccategory{Research}

\vgtcpapertype{please specify}

\title{Practices and Strategies in Responsive Thematic Map Design:\\ A Report from Design Workshops with Experts}

\author{%
\authororcid{Sarah Schöttler}{0000-0002-4898-2619},
\authororcid{Uta Hinrichs}{0000-0001-7494-0941}, and
\authororcid{Benjamin Bach}{0000-0002-9201-7744}
}

\authorfooter{
  \item
  Sarah Schöttler and Uta Hinrichs are with the University of Edinburgh.
  E-mail: \{sarah.schoettler\,$|$\,uhinrich\}@ed.ac.uk
  \item
  Benjamin Bach is with Inria and the University of Edinburgh. 
  E-mail: bbach@ed.ac.uk
}

\abstract{This paper discusses challenges and design strategies in responsive design for thematic maps in information visualization.
Thematic maps pose a number of unique challenges for responsiveness, such as inflexible aspect ratios that do not easily adapt to varying screen dimensions, or densely clustered visual elements in urban areas becoming illegible at smaller scales.
However, design guidance on how to best address these issues is currently lacking.
We conducted design sessions with eight professional designers and developers of web-based thematic maps for information visualization.
Participants were asked to redesign a given map for various screen sizes and aspect ratios and to describe their reasoning for when and how they adapted the design.
We report general observations of practitioners’ motivations, decision-making processes, and personal design frameworks.
We then derive seven challenges commonly encountered in responsive maps, and 17 strategies to address them, such as repositioning elements, segmenting the map, or using alternative visualizations.
We compile these challenges and strategies into an illustrated cheat sheet targeted at anyone designing or learning to design responsive maps.
The cheat sheet is available online: \sitelink.

}


\graphicspath{{figs/}{figures/}{pictures/}{images/}{./}} 

\usepackage{mathptmx}                  

\newcommand{\sitelink}{\href{https://responsive-vis.github.io/map-cheat-sheet}{responsive-vis.github.io/map-cheat-sheet}}

\usepackage{multirow}
\usepackage{array}
\usepackage{url}

\begin{document}



\maketitle

\section{Introduction}
\label{sec:intro}

Data visualizations, including thematic maps, are often accessed from a variety of devices such as desktop computers, smartphones, or even smart watches.
To be usable across different devices, visualizations need to be designed and developed with responsiveness in mind.
Thematic maps such as choropleth or proportional circle maps are widely used in news media, dashboards, and analytics applications to communicate data such as election results, climate data, or infectious disease prevalence to diverse audiences.
Yet, many of the generic strategies and tools developed for responsive visualization are not applicable to, or do not sufficiently address, challenges specific to thematic maps.
In particular, maps have a largely fixed aspect ratio (e.g., a map of the UK is always about twice as tall as it is wide), spatial units have varying sizes and shapes (e.g., compare small urban constituencies to larger rural ones), and data often has varying density and an uneven distribution across the map (e.g., data that is linked to population will be more dense in urban areas).
In thematic mapping, it is often desirable to keep the entire map on-screen as this will show spatial patterns and relationships most clearly.
However, rescaling maps to fit varying screen sizes may result in issues such as the (fixed) aspect ratio of the map being mismatched with the aspect ratio of the available screen space, spatial units becoming too small to be detectable, and dense urban areas becoming cluttered with many overlapping visual elements.

Practitioners are well aware of these specific challenges; for example, one practitioner interviewed by Hoffswell~et~al. referred to the US map as a \textit{``nightmare for responsiveness''}~\cite[p.8]{hoffswellTechniquesFlexibleResponsive2020} due to its landscape aspect ratio being mismatched with typical smartphone aspect ratios.
Nonetheless, reviewing collections of published responsive maps and cartograms (e.g. \cite{hoffswellTechniquesFlexibleResponsive2020,kimDesignPatternsTrade2021}) reveals that many thematic map designs primarily rely on strategies that avoid modifying the actual map, instead focusing on rearranging labels, legends, and other UI elements.
If those changes are not sufficient to address the issues, significant amounts of information are often removed, e.g., by filtering out values and removing annotations and labels.
As a consequence, generic responsiveness strategies derived from these visualization collections only address a small subset of the challenges unique to thematic maps and frequently lead to significant loss of information in the visualization.
In cartography, research on mobile and responsive mapping has largely focused on maps for navigation and wayfinding~\cite{houtmanIntersectionMobileThematic2023} as opposed to thematic maps for information visualization.
As such, there is a lack of established design solutions for digital thematic maps that balance information density with legibility on small screens.
Ultimately, this creates a worse experience for end users who may struggle to get the information they need when viewing and interacting with thematic maps on mobile devices.

In this paper, we aim to discover potential design solutions for responsive thematic maps by learning from experienced practitioners about how they create responsive maps and which strategies, solutions, and rationales they employ.
To that end, we first conducted three exploratory interviews with expert visualization and thematic map creators.
The experts highlighted the lack of best practices and tool support for responsive mapping.
The interviews also revealed external factors affecting the responsive design process, such as prioritizing designing for the clients' devices rather than the expected end users' devices, as well as trade-offs required by optimizing for responsiveness with limited time and budget.
We then followed this up with 1:1 design workshops with eight professional map visualization designers and developers with diverse professional backgrounds.
To exclude external factors, the workshops `sandboxed' the design process by asking participants to redesign a provided map for various screen sizes while discussing their process, design choices, perceived challenges, and trade-offs they were making.
Our workshops confirmed a perceived lack of established best practices, with most participants relying on their intuition and tacit knowledge.
Some also expressed insecurity about their design choices and expected other practitioners to disagree with them.
Despite this, participants' strategies, priorities, trade-offs and design choices showed significant similarities, pointing to a common set of strategies that have proven useful in practice.

Based on these insights, we first discuss general observations of practitioners' decision-making processes, personal design frameworks, and motivations, such as a common preference for thematic maps over alternative geographic visualizations, interaction being perceived as more of an opportunity than a challenge, and disagreements and difficulties related to the lack of established best practices.
We then present an overview of specific design solutions that address challenges in responsive thematic mapping.
We categorize solutions into four groups with increasingly challenging trade-offs being made: first, subtle strategies such as displacing individual elements and rearranging other UI elements; second, strategies that use scrolling, zooming, and panning to place parts of the map off-screen; third, strategies that split the map into multiple segments; and finally, strategies that replace the map with alternative visualizations.
These strategies should be applied in this order, only moving on to the next group of design solutions if previous groups have failed to sufficiently address all design challenges.
We compile these design solutions into an illustrated cheat sheet (\cref{fig:cheatsheet}) to make them available to visualization creators with less thematic mapping experience.
In a small case study with the project lead and visualization developer of a research project, they reported finding the cheat sheet a helpful guide that included many strategies they were not previously aware of.
They successfully used it to develop wireframe sketches for different screen sizes for a desktop-focused map visualization they had previously developed.

\section{Related Work}
\label{sec:rw}

While there is little prior work on responsive thematic mapping, we build on work in responsive and mobile visualization (\cref{sec:rw:responsive-vis}), as well as mobile mapping and map generalization (\cref{sec:rw:mobile-maps}).
Further, since we closely work with practitioners in this paper and ultimately aim to support their work, we briefly discuss prior work on practitioners' design processes, attitudes, and strategies (\cref{sec:rw:practitioners}).

\subsection{Responsive \& Mobile Visualization}
\label{sec:rw:responsive-vis}

\textit{Responsive visualization} is still an emerging area in information visualization~\cite{meeksResponsiveDataVisualization2014,andrewsResponsiveVisualisation2018}.
In contrast to \textit{mobile visualization}~\cite{leeMobileDataVisualization2021}, which focuses solely on mobile devices, responsive visualization is much wider in that it aims at visualization \textit{across} many types of devices.
Recent work has focused on identifying and classifying responsive visualization strategies commonly used by practitioners and developing tools and libraries to facilitate this design process for visualization creators (e.g.,~\cite{hoffswellTechniquesFlexibleResponsive2020,kimDesignPatternsTrade2021,kimCiceroDeclarativeGrammar2022}), as well as on (partly) automating this process~\cite{wuViSizerVisualizationResizing2013,wuMobileVisFixerTailoringWeb2021,kimAutomatedApproachReasoning2022}.
Much of this work has primarily focused on non-geographic visualizations such as bar charts, scatterplots, or parallel coordinate plots, as well as communication-focused contexts such as the news media.
While there are countless examples of thematic maps in data journalism, in research on responsive visualization design, maps have so far only been considered as one of many possible generic visualization types, with no special attention paid to their unique characteristics as discussed in \cref{sec:intro}.

For responsive visualization in general, Kim~et~al.~\cite{kimDesignPatternsTrade2021} introduce a taxonomy of design patterns based on a sample of 378 pairs of small and large screen visualizations collected from the media and other sources focused on communication.
Similarly, Hoffswell~et~al.~\cite{hoffswellTechniquesFlexibleResponsive2020} report on techniques for adapting visualizations for different devices obtained from 231 responsive visualizations.
Both of these collections include map and cartogram-based examples.
Hoffswell~et~al.'s collection includes 43 examples labeled as either ``map'' or ``cartogram''; Kim~et~al.'s collection includes 82 such examples.
Instead of analyzing collections of visualizations to identify design patterns and techniques, our study involves practitioners to identify challenges, design strategies, and rationales behind designs in responsive thematic mapping.

\subsection{Mobile Mapping \& Map Generalization}
\label{sec:rw:mobile-maps}

In cartography, \textit{mobile mapping} has primarily focused on maps for navigation and other reference maps, less on thematic maps for information visualization~\cite{houtmanIntersectionMobileThematic2023,leeMobileDataVisualization2021}.
Common conventions in mobile map design include strategies such as ``center map on user's location'' and ``increase size of interactive point symbols''~\cite[see Table 1]{rickerMobileMapsResponsive2018}; these and other strategies can be applied to thematic maps as well.
More generally, cartographic research has developed a variety of techniques for \textbf{map generalization}, the cartographic practice of adapting maps for different scales by ``representing various geographies at different levels of detail''~\cite[p.~1]{stanislawskiGeneralisationOperators2014}.
There are a variety of generalization techniques, such as line simplification algorithms, automated displacement of elements or labels, or strategies such as typification or symbolization.
In cartography, these strategies are known as \textit{generalization operators}~\cite{rothTypologyOperatorsMaintaining2011}.
Similar to mobile mapping, generalization research has largely focused on topographic and general-purpose maps---as opposed to thematic maps---but there has recently been a push to consider generalization operators for thematic maps as well~\cite{raposoChangeThemeRole2020}.
Despite the seemingly opposing goals---responsive visualization aims to maintain the same message across all sizes of the visualization, whereas generalization aims to show information relevant to the scale---many generalization techniques can be useful in the context of responsive geographic visualization.
This includes techniques such as such as line simplification, symbolization, and typification~\cite{rothTypologyOperatorsMaintaining2011}.
Setlur and Chung~\cite{setlurSemanticResizingCharts2021} even argue that generalization techniques could be applied to non-geographic charts as well and repurpose a line simplification technique for responsive line charts.
In our workshops with practitioners, we saw several techniques drawn from approaches in generalization, such as adjusting the scale of symbols, reducing outlines, or using schematized map designs.

\subsection{Practitioners' Design Processes}
\label{sec:rw:practitioners}

Besides best practices and techniques, we are interested in the reasoning applied by practitioners and their decisions in responsive map design in this work.
Generally, iteration, exploration, and tacit experiential knowledge are crucial to visualization design processes.
Guidelines (e.g., \cite{diehlVisGuidesForumDiscussing2018}) and design methodologies (e.g., \cite{sedlmairDesignStudyMethodology2012, munznerNestedModelVisualization2009}) can support these processes but are not strictly followed by practitioners: interviewing 20 visualization designers about their respective design practices, Parsons~\cite{parsonsUnderstandingDataVisualization2022} reveals several aspects of how visualization practitioners work.
For example, designers employ a wide variety of approaches without consensus on specific steps or a systematic process.
Rather, visualization design involves organic and intuitive decision-making, iteration, and `planning in action'.
Mostly, knowledge is tacit, subjective and of experiential nature.
High-level theories (such as information theory) are rarely used, although guidelines and empirical findings from academic papers can play a role, for example \textit{Gestalt Laws} or findings on \textit{chart junk}~\cite{batemanUsefulJunkEffects2010}.
Practitioners prefer frameworks that do not prescribe but support reflection and decision-making~\cite{stoltermanNatureDesignPractice2008}.
Defining clear tasks and constraints is key.
Evaluation of work is then mostly subjective or through colleagues, rather than through rigorous and structured evaluation methodologies, and many of the interviewees in Parsons' study did not express specific rationales for deciding on, e.g., a visual representation.
These findings reflect earlier work by Schön in the \textit{Reflective Practitioner}~\cite{schonReflectivePractitionerHow2017}.

Our study confirms many of these previous findings: even in our relatively structured, `sandboxed' design study, designers worked ad-hoc and made decisions intuitively and iteratively.
Yet, when asked, participants were able to articulate clear rationales, trade-offs, and design considerations in responsive map design, allowing us to derive general strategies.
We hope our findings and strategies help experienced designers to think more deliberately about responsive thematic map design by providing structured guidance and theory.
For novices, in addition to the above, we hope to provide practical, actionable design suggestions and help them understand considerations and decisions involved in the design process.

\section{Study Methodology}

To identify challenges and strategies in responsive thematic map design, we initially conducted three preliminary expert interviews (\cref{sec:interviews}).
These interviews revealed many practical hindrances when designing responsive thematic maps, but also pointed to a variety of techniques used by experts.
Informed by these interviews, we then developed a 1:1 design workshop format, which we ran with eight professional map and visualization designers and developers (\cref{sec:workshops}) to identify more specific challenges and design solutions.

\begin{figure*}[t]
	\centering
	\includegraphics[height=7cm]{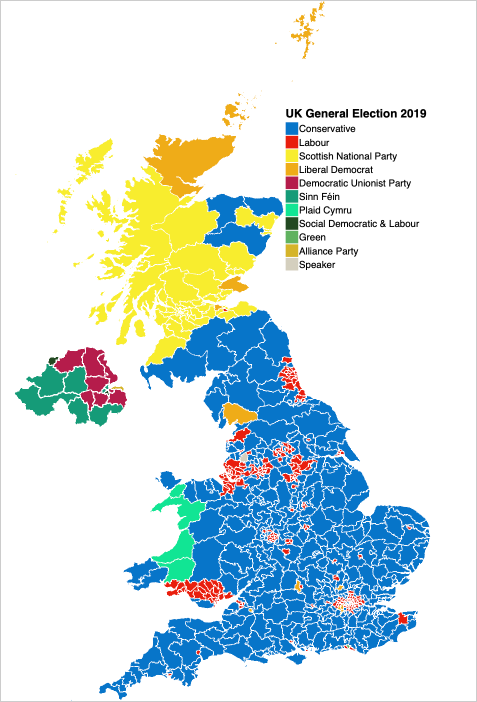}
	\hfill
	\includegraphics[height=7cm]{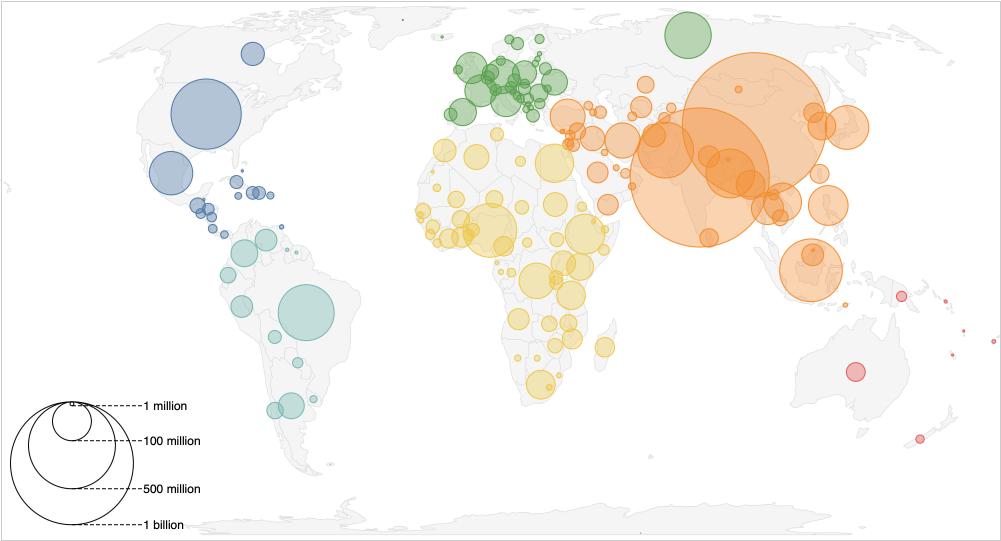}
	\caption{Workshop participants were randomly allocated one of these two maps: a choropleth map of the 2019 General Election results in the UK (left) and a world map showing population by country (right).}
	\label{fig:maps}
\end{figure*}

\subsection{Preliminary Expert Interviews}
\label{sec:interviews}

To inform our main study design (\cref{sec:workshops}), we conducted three preliminary expert interviews.
Prior work~\cite{hoffswellTechniquesFlexibleResponsive2020} had reported on practitioners' design practices for responsive visualization in general, but did not discuss maps in detail.
Thus, our goal in these interviews was to better understand differences between responsive thematic mapping and generic responsive visualization:
\textit{%
	What challenges and trade-offs unique to thematic maps do practitioners encounter?
	To what extent are thematic maps supported by tools and techniques for responsive visualization?
	How do design processes for responsive thematic maps differ from other responsive visualizations?
}
To answer these questions, we conducted three semi-structured, hour-long interviews with expert visualization designers and developers.

\textbf{Participants---}%
We interviewed three experts (E1--E3), who each had between 10 and 20 years of experience in interactive visualization and map design and development.
E1 primarily works on interactive maps; E2 and E3 work on a variety of visualization projects but had recently completed major map-based projects.
E1 is primarily a developer, E2 does development as well as design work, and E3 is primarily a designer.

\textbf{Interview Setup---}%
Participants were interviewed individually in hour-long online calls, where we recorded video and audio.
In the first segment of the interview, we asked general questions about their experience with map visualizations and responsiveness.
We then asked them to describe the design process for a recent responsive map-based project they had worked on in detail, covering the different steps in their design process, major challenges, tool usage, and detailed explanations of how responsiveness was designed for in the project.
Participants shared their screen for this part.
Finally, we asked about their suggestions for improving design support for responsiveness in visualization, and how they would change their design choices if resource constraints were irrelevant.

\textbf{Results---}%
All three of the experts primarily rely on \textbf{generic responsive web design guidance and tooling} but switched between multiple tools and used workarounds and plugins to adapt tools to the specific needs of responsive map and visualization design.
E3 highlighted performance issues when working with maps: For example, \textit{Figma}, a popular interface and UX design tool, gets sluggish when complex vector data---such as country outlines---is imported, hampering the map design process.
Yet, all three experts acknowledged that technological advancements for responsive web design in recent years had made many aspects of responsive visualization design easier as well.
The three experts were also in agreement on the main \textbf{factors to consider in responsive visualization}, with screen size and aspect ratio being the main challenges.
Differences between interaction modalities mainly require accounting for the lack of mouseover interaction on touchscreen devices.
Finally, all experts were mindful of limiting file sizes and computational complexity for mobile devices, which may be less performant and have slower or less stable internet connections.
All of the interviewees referenced \textbf{map-specific challenges}: for example, fixed aspect ratios make world maps difficult to show on portrait-format mobile screens, and differently sized spatial units lead to small countries being difficult to visibly show on a world map.
They each shared \textbf{design solutions} they had considered or implemented, such as removing county outlines on small maps, using different map projections, or making use of interactive features ranging from panning and zooming to custom selection interactions to address the \textit{``fat-finger''} problem.
E2 and E3 generally try to keep functionality similar across versions, whereas E1 regularly removes less important features for mobile users.
\textbf{Design decisions} were described as iterative and heavily contextual, depending on stakeholders/clients, anticipated usage scenarios, and specific characteristics of the visualized data.

\subsection{Design Workshops with Thematic Map Creators}
\label{sec:workshops}

\begin{figure*}[t]
	\centering
	\includegraphics[width=0.49\linewidth]{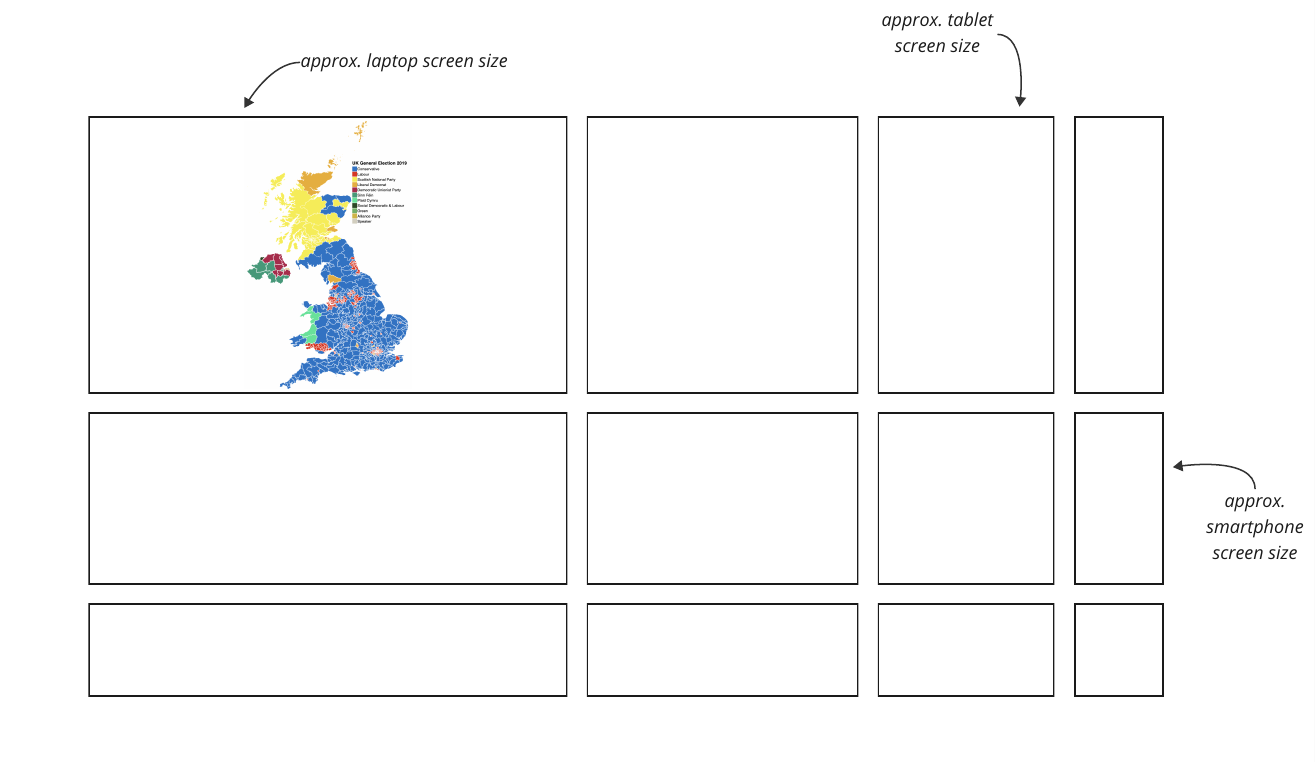}
	\hfill
	\includegraphics[width=0.49\linewidth]{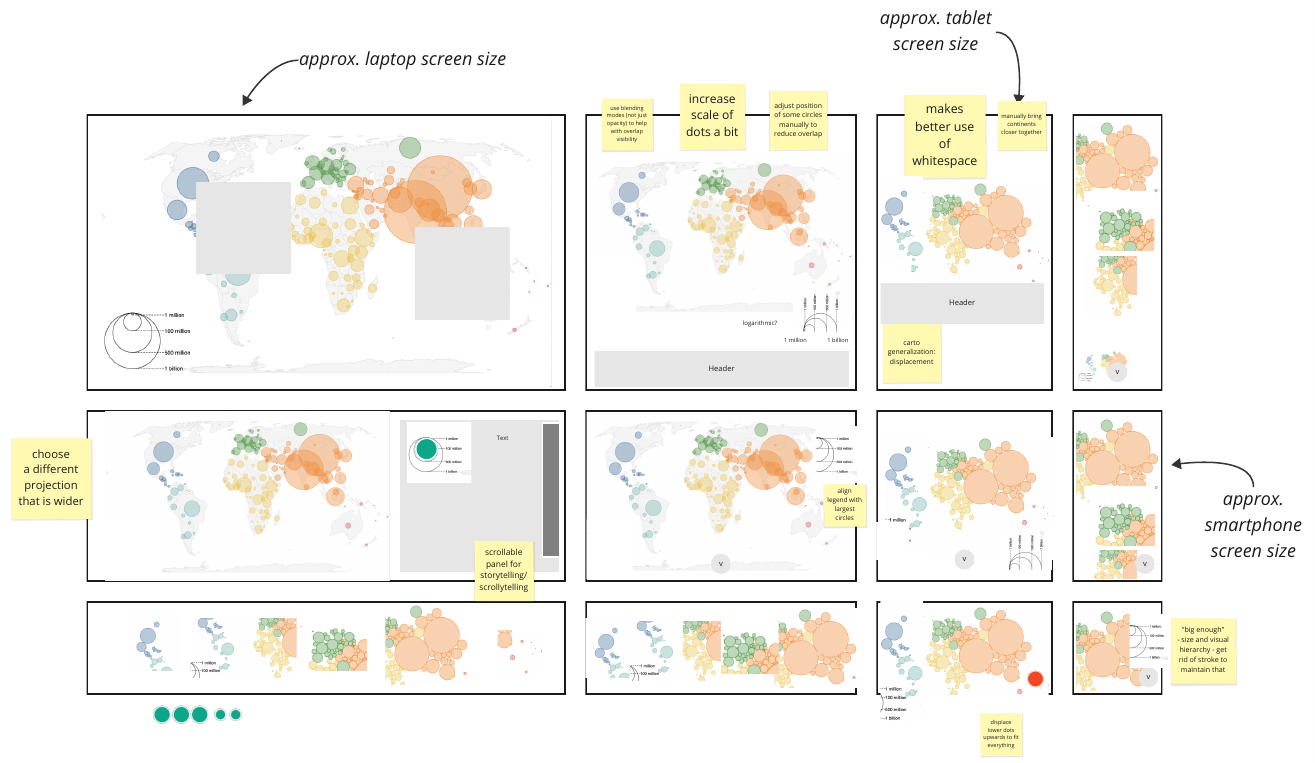}
	\includegraphics[width=0.49\linewidth]{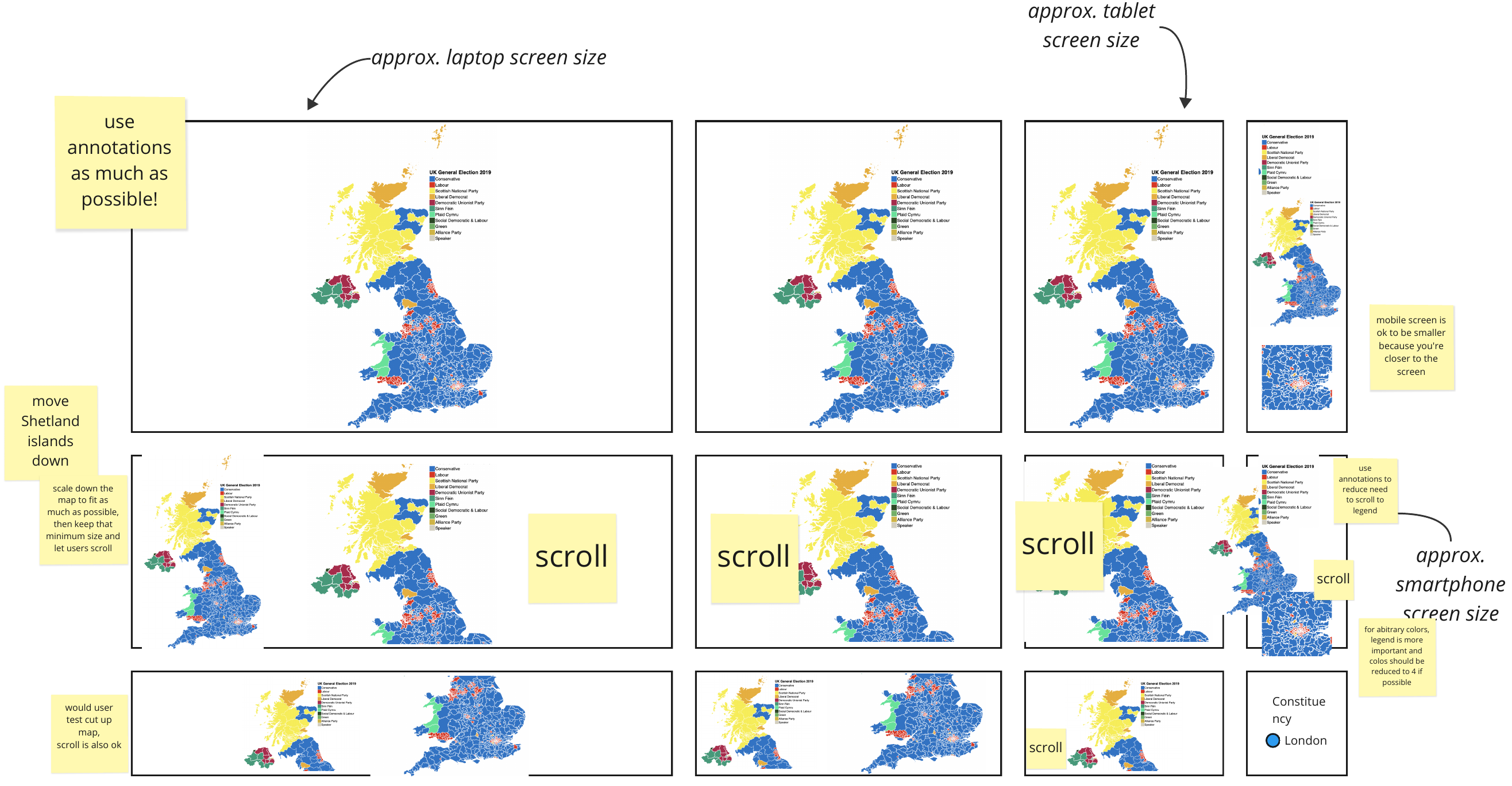}
	\hfill
	\includegraphics[width=0.49\linewidth]{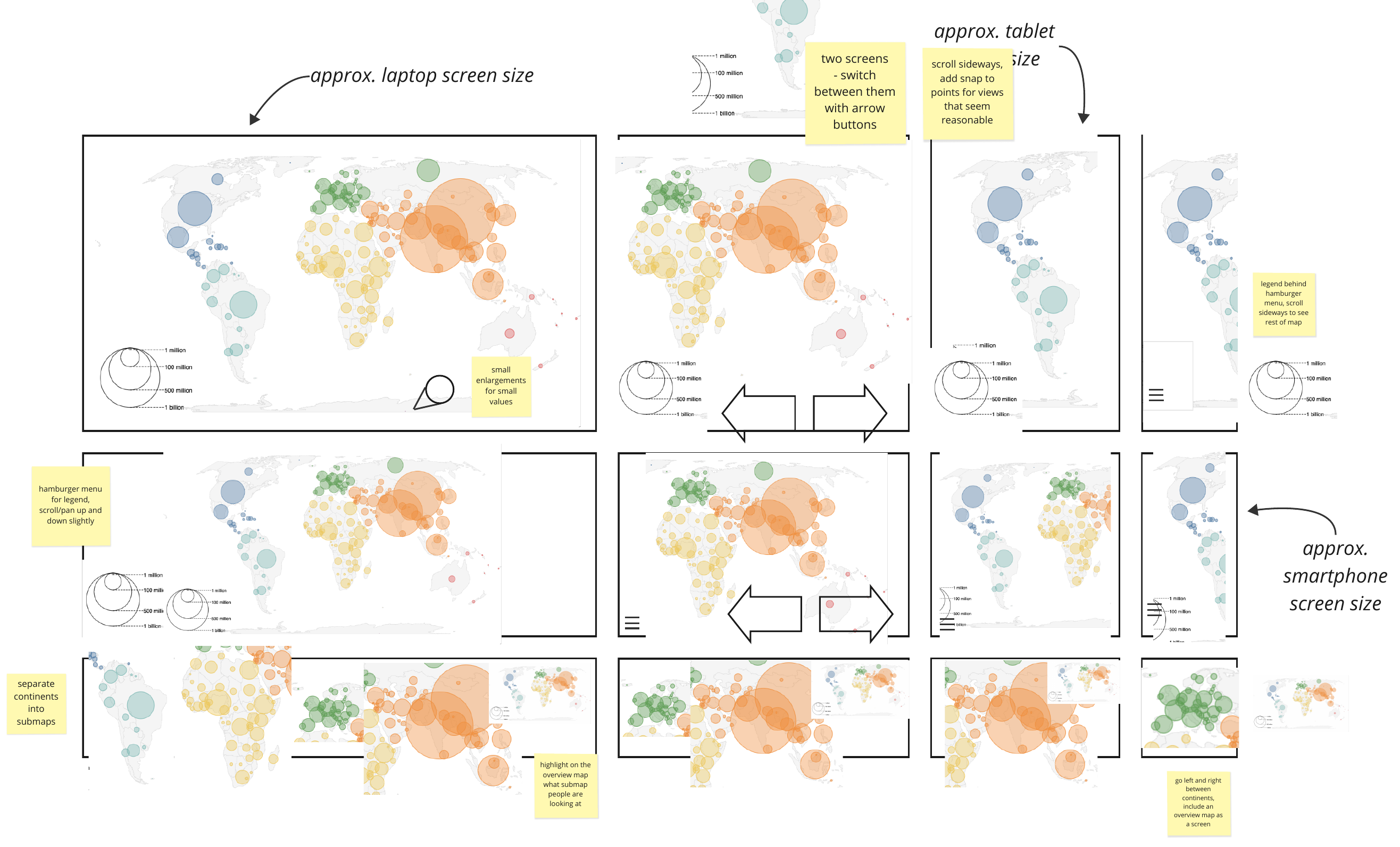}
	\caption{Excerpts from the Miro board used for the study.
		Participants were provided with one of two maps, a choropleth map of UK election results (left) or a proportional circle map showing global population by country (right).
		The top left image shows the empty Miro board provided to each participant in the beginning of the workshop.
		Some of the rectangles are labeled with the common devices they correspond to (laptop, tablet, smartphone); it was explained to participants that the remainder were in-between sizes.
		Participants created designs by repositioning, resizing, duplicating, and cropping the original map, as well as adding additional elements and images found on the internet, supplemented with (virtual) sticky notes to specify designs.
		We show the final Miro boards of three participants here: P6 (top right), P2 (bottom left), and P5 (bottom right).
		Full-size versions of these Miro boards excerpts are available in the supplemental material.
	}
	\label{fig:miro}
\end{figure*}

Motivated by these insights, we conducted eight 1:1 workshops with experienced thematic map designers and developers to identify specific design solutions that could serve as practical guidance for designers.
We developed our workshop format to exclude the many practical hindrances described in our preliminary interviews, such as needing to rely on workarounds and switching between tools, as well as time and budget pressures.
Instead, we engage participants in a \textit{`sandboxed'} design process with a focus on design solutions and rationales rather than implementation.

\textbf{Participants---}%
We recruited eight professional visualization designers and developers via personal contacts, conferences, and online.
Participants had to have at least two years of full-time experience (or equivalent part-time/freelance experience) in visualization design and/or development, some of which needed to be with maps.
We were aiming to recruit participants with diverse professional backgrounds and in different sectors.
Of our eight participants, three participants worked in government organizations, two in research institutions/organizations, two in journalism, and one in finance.
All participants had experience in visualization design as well as development; three participants additionally had significant visualization research experience, one of whom was also a trained cartographer.
Two participants were female, the remainder male.
One workshop was conducted in person, where we recorded audio only, the remainder online via video call, where we recorded both video and audio.
None of the participants were involved in the preliminary expert interviews described in the previous subsection.

\textbf{Map Selection---}%
Participants were randomly provided with one of two maps in our workshops:
a choropleth map showing the 2019 UK General Election results, or a proportional circle map showing global population by country (\cref{fig:maps}).
These maps were chosen to represent two different aspect ratios (portrait and landscape formats), as well as the most common types of thematic maps.
Most thematic maps (e.g., see~\cite{hoffswellTechniquesFlexibleResponsive2020,kimDesignPatternsTrade2021} for collections from major news outlets) fall into one of two categories:
\textit{1)} choropleth and similar maps where data is encoded by shading spatial units in the map, or \textit{2)} maps consisting of a basemap with an overlaid symbol layer, such as dots, proportional symbols, markers, or isolines.

\textbf{Workshop Setup---}%
We provided participants with a link to a Miro board containing the map and twelve rectangles of various sizes and aspect ratios as shown in \cref{fig:miro}.
Some rectangles represented common screen sizes: a laptop screen (largest rectangle), a tablet screen in two orientations, and a smartphone screen in two orientations.
The other rectangles are not commonly-used screen sizes, but we included them for two reasons: many of these sizes may occur when, e.g., a user has an on-screen keyboard open, resizes their browser window, or some of the screen space is covered by other UI elements that cannot be moved.
More importantly, we wanted to include the in-between sizes to encourage participants to gradually adjust designs and identify the exact points at which a design breaks.
We opted for a Miro board over any more specialized visualization or map design tools to generate design strategies that are as tool-agnostic as possible.
The initial map was placed in the largest rectangle and scaled to fit.
This reflects the most common approach in practice: our expert interviews as well as prior work~\cite{hoffswellTechniquesFlexibleResponsive2020} indicate that visualization practitioners usually design a desktop-size visualization first, then adapt it for other formats.

\textbf{Procedure---}%
An introduction instructed participants to purely focus on the map design without worrying about practical implementation challenges.
We did not provide participants with additional context or a specific target group for the provided map, instead encouraging them to project their own context drawn from their experience onto the design exercise.
We also did not specify what aspects of responsiveness participants should consider---while our Miro board rectangle setup was focused on differently sized screens, they were free to discuss additional factors.
We then asked participants to create map visualization designs for each of the twelve screen sizes represented by rectangles.
To create design ideas, participants could use a free-form sketching tool, use basic shapes and text boxes, copy-paste images from the internet, and rearrange, crop, and duplicate existing images.
While creating designs, participants were asked to voice their design choices, explain considerations, and discuss the advantages and disadvantages of their various designs.
At times, the interviewer would ask follow-up questions to better understand some explanations.
At the end of the workshop, interviewer and participant reviewed all twelve designs, filled in any gaps, made designs more consistent, and clarified any open questions.

\textbf{Pilot Study---}%
Prior to recruiting participants, we piloted the workshop with two early-career visualization developers to ensure instructions were clear and the timeline feasible.
This led to some changes in the Miro board, notably the introduction of the rectangles to provide more structure to the design, and the removal of a gallery of example visualizations.
We had initially included this to provide some inspiration and streamline the sketching process (reducing the need to find images online), however, our pilot participants felt this gallery---showing various maps, cartograms, and other visualizations---biased their design choices.

\section{Design Strategies for Responsive Maps}

\newcommand{\includebox}[1]{
	\hfill\includegraphics[width=0.98\columnwidth]{img/box_#1.pdf}\hfill
	\vspace{0.15cm}
}

We first report on some general observations from the workshops, then on the specific challenges and design solutions derived from the workshops.
\Cref{sec:cheatsheet} then introduces our illustrated cheat sheet which summarizes these challenges and solutions.

\begin{table}
	\setlength\tabcolsep{0.1cm}
	\renewcommand{\arraystretch}{1.2}
	\centering
	{\footnotesize
		\begin{tabular}{l|>{\raggedright\arraybackslash}m{3.8cm}|cccc|cccc}
			     &                                                                & \multicolumn{4}{c|}{\textit{UK Election}} & \multicolumn{4}{c}{\textit{Population Map}}                                                                                     \\
			     & \textbf{Design Solution}                                       & \textbf{P1}                               & \textbf{P2}                                 & \textbf{P3} & \textbf{P4} & \textbf{P5} & \textbf{P6} & \textbf{P7} & \textbf{P8} \\
			\hline\hline
			S1.1 & reposition UI elements                                         & $\times$                                  & $\times$                                    & $\times$    & $\times$    & $\times$    & $\times$    &             & $\times$    \\
			S1.2 & re-design legend                                               &                                           &                                             & $\times$    &             & $\times$    & $\times$    &             &             \\
			S1.3 & remove or hide legend                                          & $\times$                                  & $\times$                                    & $\times$    & $\times$    & $\times$    &             & $\times$    &             \\
			S1.4 & adjust outline widths \& fill color, opacity \& blending modes &                                           & $\times$                                    &             &             &             & $\times$    &             &             \\
			S1.5 & adjust scale of symbols                                        &                                           &                                             &             &             &             & $\times$    &             &             \\
			S1.6 & change map projection                                          &                                           &                                             &             &             &             & $\times$    &             &             \\
			S1.7 & displace symbols                                               &                                           & $\times$                                    &             &             &             & $\times$    &             & $\times$    \\\hline
			S2.1 & vertical scroll                                                &                                           & $\times$                                    &             &             &             &             &             &             \\
			S2.2 & horizontal scroll                                              &                                           &                                             &             &             & $\times$    &             & $\times$    &             \\
			S2.3 & pan and zoom                                                   &                                           &                                             & $\times$    & $\times$    & $\times$    &             &             & $\times$    \\
			S2.4 & zoom cutout or inset                                           &                                           & $\times$                                    &             &             & $\times$    &             &             & $\times$    \\\hline
			S3.1 & segment map -- straight line                                   &                                           & $\times$                                    &             &             & $\times$    &             &             & $\times$    \\
			S3.2 & segment map -- geographic units                                & $\times$                                  &                                             & $\times$    &             & $\times$    & $\times$    &             &             \\ \hline
			S4.1 & cartograms/grid maps                                           & $\times$                                  &                                             &             &             &             & $\times$    & $\times$    &             \\
			S4.2 & partly geographic visualizations                               & $\times$                                  &                                             &             &             &             & $\times$    &             &             \\
			S4.3 & non-geographic visualization                                   &                                           &                                             &             &             &             &             & $\times$    &             \\
			S4.4 & lookup box or table (no visualization)                         &                                           & $\times$                                    &             &             &             &             & $\times$    &
		\end{tabular}
		\caption{Overview of design solutions identified in the design workshops and which participants used or recommended them.
			This list follows the same structure as the cheat sheet (Fig.~\ref{fig:cheatsheet}).
		}
		\label{tab:overview}
	}
\end{table}

\subsection{General Observations}

\textbf{Maps are preferred by designers (and users)---}%
Most participants expressed wanting to preserve a map-based visualization across screen sizes whenever possible, citing reasons such as maps being familiar to users, good at showing geographic patterns, their aesthetics and visual impact, and that users tend to like them.
P7 challenged this view somewhat, saying that
\textit{%
	``if there is no geographical pattern [\dots] I would try to find [a different visualization], because maps are quite tricky to make responsive.
	It's not that I hate maps, but often we overuse them.''%
}
The concept designs created in the workshops reflect this preference: all participants used maps for a majority of the screen sizes, four participants even used maps across all screen sizes.
Switching to alternative visualization designs was generally treated as a last resort, and three of the four participants who did opted for visualizations that preserve some geographic context, such as cartograms, grid maps, or spatially ordered treemaps.
In our overview of design solutions (\cref{sec:solutions}), we reflect this preference by grouping design solutions into four groups S1--S4, where solutions in S1 fully preserve the map, S2 and S3 solutions modify the map more significantly, and S4 solutions remove the map.

\textbf{Lack of established design guidance---}%
Many participants expressed insecurity about their design choices, implying a lack of established design guidance in this area to fall back on.
Despite this, all participants were able to justify their choices clearly and some had significant experience to draw on.
Additionally, participants mostly used techniques that were similarly or identically used by other participants, indicating that, despite its lack of formalization, there exists a set of techniques that have proven useful in practice---this is what we aim to provide an overview of in \cref{sec:solutions}.

\textbf{Interaction as opportunity---}%
Interaction is often framed as a challenge in responsiveness, requiring interactive features to be adapted to various devices~\cite{horakResponsiveVisualizationDesign2021}.
Our participants generally perceived interaction as more of an opportunity than a challenge: with the exception of mouseover labels, browsers nowadays automatically deal with the differences in input devices, meaning that developers rarely need to manually adjust this.
Interaction was frequently used to solve for a lack of space, for example by making use of scrolling, zooming, and panning when a map cannot fully fit in the available screen space, or by hiding the legend in a context menu to save on screen space.

\textbf{Disagreement on challenges---}%
While most participants were in agreement that fixed aspect ratios as well as small and overlapping elements are major issues in responsive geographic visualization, one participant notably disagreed with this: P4 felt that enabling pan and zoom (as is standard in many web-mapping libraries) solves any issues regarding mismatched aspect ratios or small areas/elements sufficiently.
In the preliminary expert interviews, E1 shared a similar view, stating that \textit{``the map is the easy part''} and that their designs mainly relied on rearranging other UI elements.
Similarly, P4 relied purely on pan and zoom alongside layout changes for all screen sizes in the workshop.
The remainder of the participants were not satisfied with this approach: they cited motivations such as not wanting to lose the \textit{``impact''} (P3) of the map, users needing to see the full map to be able to compare areas and see the overall distribution of data, and users often not engaging with interactive features.

\subsection{Challenges}

We now discuss challenges in responsive map design, as identified by our workshop participants.
Some of these challenges have previously been discussed~\cite{hoffswellTechniquesFlexibleResponsive2020} and may appear obvious.
However, we wanted to ensure that \textit{a)} we did not miss any hidden additional challenges and that \textit{b)} we do not consider challenges that do not exist in practice.
We also felt it useful to provide this overview for those less experienced with maps, particularly in the context of the cheat sheet presented in \cref{sec:cheatsheet}.
Additionally, our participants shared some detailed insights on how to assess to what extent these challenges apply to a given map.
Notably, text labeling was not seen as a major challenge for thematic maps in our expert interviews nor our workshop study---despite being a common issue in mobile maps outside of information visualization, with many techniques proposed to automate label placement~\cite{bekosExternalLabelingTechniques2019}.
This may be due to fewer labels being required on thematic maps, or interactive tooltip labeling being more common in thematic mapping.
We identify three types of challenges: \textbf{parts of the map becoming too small} (C1), \textbf{mismatched aspect ratios} (C2), and \textbf{overlapping legends and other UI elements} (C3).
Design solutions to address these are discussed in the following subsection.

\subsubsection{C1 --- Parts of the map become too small}

\includebox{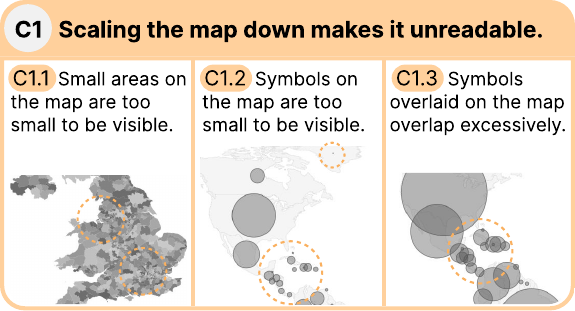}

Maps, and the data displayed on them, have varying densities throughout.
For example, densely populated areas will in many contexts have smaller spatial units and/or more data points associated with them than sparsely populated rural areas.
As the map gets scaled down to fit smaller screen spaces, this leads to three main challenges:
First, small spatial units (e.g., electoral constituencies in central London) become too small to be clicked and/or too small to be visible (C1.1).
Particularly in choropleth maps, where the data is encoded in the fill color of spatial units, this results in parts of the data being lost from the visualization, which all participants working on the choropleth map (P1--P4) recognized.
Where the map is only acting as a basemap (and the data is encoded as a symbol layer on top), this poses less of a concern.
For example, P8 suggested that unless one of the goals of the map is to communicate something about the small area in question, it does not matter if it disappears, particularly when users have access to search or zoom functionality.

Second, symbols overlaid on the map may become too small (C1.2).
Depending on context, `too small' may refer to the size needed for users to reliably tap or click an element~\cite{siekFatFingerWorries2005}, or it may refer to the size needed to visually detect the element~\cite{clevelandModelStudyingDisplay1993}.
For example, the smallest circles in a proportional circle map might become too small to be detectable, a challenge experienced by all participants working with the population map (P5-P8).
This challenge can be mitigated by scaling up the visual elements to make them visible once more, however, this will exacerbate any overlap between elements (C1.3).
As such, there is always a trade-off between these two challenges: scaling symbols up helps their visibility at the cost of increased overlap, whereas scaling symbols down reduces overlap at the cost of their visibility.
P1 additionally noted that some types of color vision deficiencies may make it difficult to identify colors in small elements.
We did not ask participants to quantify thresholds for size or overlap.
A prior research study that aimed to quantify minimum legible sizes for different types of symbols on digital maps for different screen resolutions makes some tentative recommendations, but was unable to establish definitive thresholds~\cite{ledermannMinimumDimensionsCartographic2023}.

\subsubsection{C2 --- Mismatched aspect ratios}

\includebox{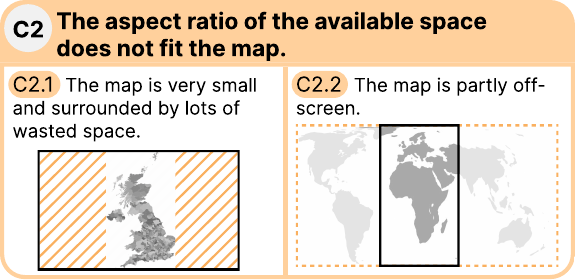}

Maps have largely fixed aspect ratios, e.g., a map of the UK will always be about twice as high as it is wide.
Depending on how much the map is scaled down to fit a different aspect ratio, this leads to one, or both, of the following issues:
If the map is scaled down to fit the screen, it will be surrounded by white space (C2.1).
Participants generally considered this an inefficient use of space, and found that, aesthetically, it created \textit{``imbalanced designs''} (P8) and a loss of \textit{``impact''} (P3).
One solution that several participants used is to use the empty space for something else, e.g., by rearranging other UI elements or adding an additional chart or explanatory text into the empty space.
However, this works only while the map is not too small (C1) when scaled to fit.
It may also lead to repeated layout changes for different screen sizes and aspect ratios, which is time-intensive and more error-prone from a development perspective.

Alternatively, the map could be scaled up so that it is partly off-screen (C2.2).
This resolves issues related to the size of the map but is often undesirable since it makes it harder to see patterns in the map as a whole.
Some solutions for this accept that the map is partly off-screen and let users scroll (P2, P5, P7) or pan and zoom (P3, P4, P5, P8) to access the rest of the map.
Alternative solutions are discussed in the following \cref{sec:solutions}.

\subsubsection{C3 --- Cannot fit the legend or other UI elements}

\includebox{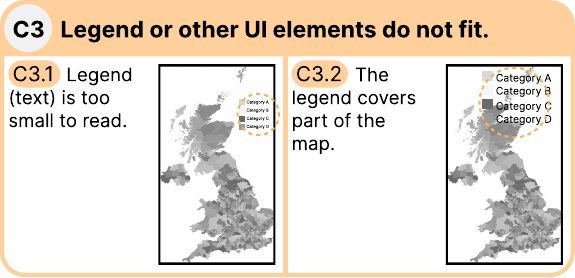}

Finally, many maps have legends or other UI elements, such as zoom or filter controls.
Since both of the maps we provided our participants with had legends, all participants recognized that this leads to one of two challenges:
if legends, or other UI elements, are scaled down with the map, the text used in them will become too small to read (C3.1).
Alternatively, if they are scaled up, they will eventually overlap the map and take up valuable space that is required for the map itself (C3.2).
Participants used various solutions for this, most commonly repositioning UI elements (S1.1) or removing or hiding the legend (S1.3).
Three participants chose to re-design the legend.
P8 additionally suggested the legend could also act as a filtering interface for the map, combining two UI elements into one.

\subsection{Design Solutions}
\label{sec:solutions}

Our workshop participants used a variety of design solutions to mitigate the challenges described above.
The overarching strategy was remarkably similar across all participants: they mostly wanted to preserve a fully visible map as much as they could, leading them to start with a variety of \textbf{subtle changes} that would allow it to scale down further while maintaining readability (S1).
Next, they moved on to strategies that compromise on this by maintaining the map but using \textbf{scrolling, panning, and zooming} to allow them to make it very small or place it partly off-screen (S2).
Another strategy is to separate the map into several \textbf{segments} that can then be rearranged to fit the screen (S3).
Only as a last resort, when these compromise strategies failed, they moved on to using \textbf{alternative visualization types} (S4).
\Cref{tab:overview} provides an overview of all design solutions and which participants used or recommended them during the workshop.
All solutions are also illustrated in our cheat sheet (\cref{fig:cheatsheet}).
The solutions we propose are \textit{tool-agnostic}, although not all solutions will be possible to achieve with all tools.
There is no 1:1 link between the challenges and design solutions described here; rather, the more challenges and the more severe they are, the more advanced the required design solutions will be.
Thus, for minor challenges, solutions from groups S1 or S2 may be sufficient, whereas solutions from S3 or S4 may be required for more severe design challenges.

\subsubsection{S1 --- Subtle design changes}

\includebox{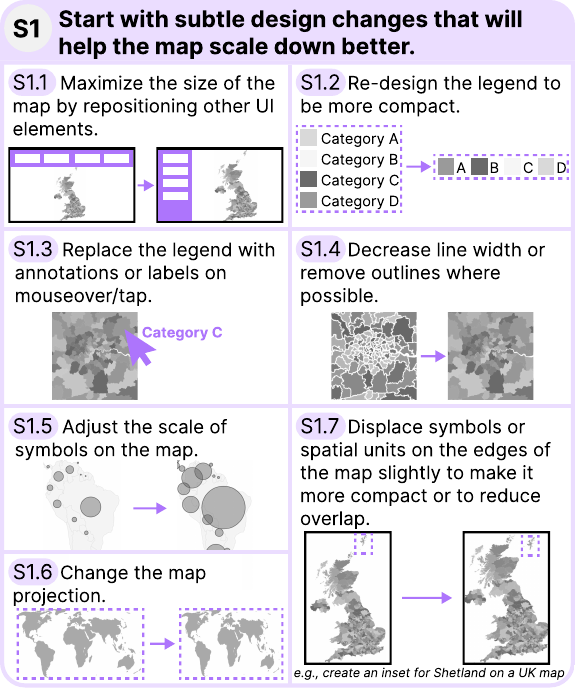}

First, participants aimed to maximize the size of the map by \textbf{repositioning other UI elements} (S1.1).
All but one participant repositioned the legend to fit around the map, and those who had introduced additional UI elements repositioned these frequently too.
As a relatively generic strategy, this design solution is also included in Kim~et~al.'s design patterns~\cite{kimDesignPatternsTrade2021}.
Second, three of the participants suggested \textbf{redesigning the legend} to be more compact, integrating it better with the map to be more space-efficient (S1.2).
For example, P6 cropped the circle legend in the population map (\cref{fig:miro}--right) into a semicircle and placed it close to some of the larger circles on the map.
Third, six participants suggested removing the legend entirely and replacing it with annotations or labels that are displayed on mouseover/tap, or hiding the legend and only showing it when the user clicks a button (S1.3).
Fourth, two participants suggested small changes to colors and outlines (S1.4): making the outlines of spatial units thinner or removing them where possible in order to maximize the fill area of small spatial units and symbols (P2), and using opacity and blending modes to make overlapping elements more visible (P6).
Only one participant adjusted the scale of symbols on the map (S1.5), however, only four of the participants were working with a map that had overlaid circles.
Symbol size can be increased to improve their visibility, as done by P6 (C1.2), or reduced to alleviate overlap (C1.3).
A sixth design solution is the use of a different map projection (S1.6), which can slightly adjust the aspect ratio of small-scale maps.
While some participants discussed this as an option, only one participant (P6) made use of this.
This solution is limited in that it only works for maps of large areas (such as an entire continent or the entire world), and choosing a map projection already requires trade-offs between distorted shapes, area sizes, and distances, which may not always make this a viable option.

Finally, three participants suggested displacing symbols or parts of the map (S1.7), a common strategy in map generalization~\cite{stanislawskiGeneralisationOperators2014}.
Displacing elements at the borders of the map---such as P2, who created an inset for the Shetland islands in the far North of Scotland, can make it more compact, allowing it to be slightly scaled up.
Alternatively, displacement can be used to alleviate overlap in particularly dense areas, as was done by P6 and P8.
Notably, despite being an established cartographic technique, displacement is poorly supported by tools and therefore challenging and time-intensive to implement in web-based maps.

\subsubsection{S2 --- Scrolling, zooming, and panning}

\includebox{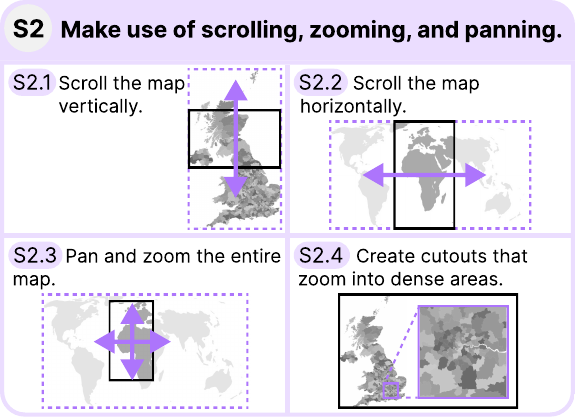}

Once these more subtle strategies were exhausted, participants started relying on user interaction more.
A common strategy was to maximize either the map’s width or its height to fit the screen, and then rely on vertical scrolling (S2.1) or horizontal scrolling (S2.2) respectively to access the rest of the map.
Letting users freely pan and zoom the entire map (S2.3) was a less popular strategy, despite being the default setting in many web mapping libraries.
In the context of scrollytelling or other web pages with significant scroll height, P7 described their reasoning for this preference as follows:
\textit{%
``[Scrolling] feels quite natural to the user, [but] I usually try to avoid panning because it kind of breaks the flow of the page.''
}
P5, on the other hand, used horizontal scroll and suggested adding several \textit{``snap points''} focused on specific areas of the map, depending on the areas particularly relevant to the story.
A similar design focused on pan and zoom was used by both P7 and P8, who suggested labeled buttons that zoom to specific areas of the map.
Finally, three participants used a (small) map scaled to fit the screen and supplemented this with one or more cutouts or insets zoomed into the densest area(s) of the map (S2.4).
For example, P2 created a zoomed in cutout of London, the densest area in the UK map.
However, P2 said they would limit the number of cutouts to 2-3, so this solution is only viable if the number of particularly dense areas in the map is relatively limited.

\subsubsection{S3 --- Segmenting the map}

\includebox{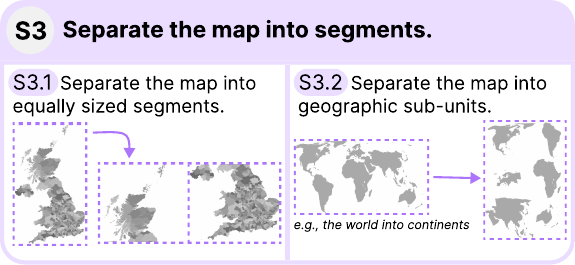}

Next, participants created various designs that rely on segmenting the map.
There are two overarching strategies: first, the map can be cut along straight lines into equally sized, rectangular segments (S3.1).
There was some disagreement on how many segments is appropriate as a maximum: P2 suggested cutting the map into no more than two segments, but other participants used up to four segments.
Alternatively, the map can be segmented into geographic sub-units (S3.2) where this makes sense in context.
Two participants each segmented the world map into continents, and the UK map into its four constituent nations.
All participants who employed this strategy emphasized that the segments should be arranged either loosely following their geographic arrangement, e.g., arranging continents West to East from left to right, or following another order that makes sense in the context of the area being mapped.
Designers only used these strategies for extremely mismatched aspect ratios and very small screens, and only when most strategies described previously had failed to resolve the challenges.

\subsubsection{S4 --- Alternative visualization types}

\includebox{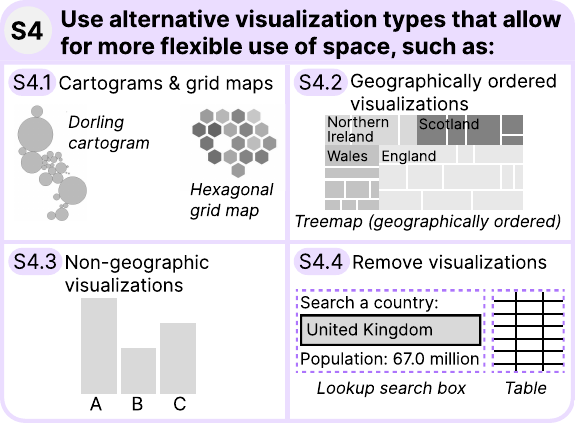}

As a last resort, once all other strategies S1-S3 could not solve the problem, participants removed the map and replaced it with a different visual representation of the data.
Three participants used cartograms (S4.1); all of them opted for Dorling cartograms or similar designs.
A Dorling cartogram is a commonly used area cartogram~\cite{dorlingAreaCartogramsTheir1996,dentCartogramValuebyareaMapping2009}, where spatial units (e.g., countries) are represented by circles proportionally scaled based on data; the circles are then geographically arranged as much as possible while avoiding any overlap between circles.
Other participants, particularly P2, were opposed to cartograms, suggesting users would struggle to understand them.
P6, on the other hand, suggested that news outlets in particular can train their readership to understand more unusual visualizations such as cartograms.
P6 shared an experience where one newspaper rejected a cartogram they had created---claiming their readers would not understand it---whereas another newspaper accepted it, having used cartograms frequently in the past.
P1 and P6 used several \textit{partly} geographic visualizations (S4.2): both participants suggested treemaps, which are space-filling and therefore extremely space-efficient but can still be approximately geographically arranged.
With a similar justification, P1 opted for a beeswarm plot, which allows for geographically ordering dots along its axis.
Finally, two participants used either non-geographic visualizations (S4.3), such as bar charts (P7), or represented the data entirely without visualizations (S4.4) using a table (P7) or a lookup search box (P2).

A potential solution that is notably not included here is \textit{rotating the map}.
While this has been suggested in prior responsive visualization research~\cite{hoffswellTechniquesFlexibleResponsive2020}, participants in our study generally did not consider this a suitable solution, which is why it is not included here.
Although there is no user study that has formally evaluated this approach, P8 said
\textit{%
	``I wouldn't try to rotate the [map] because then the person rotates the phone and then [the map] rotates again.''
}
Similarly, in our preliminary expert interviews, E3 described considering a rotated map design for mobile screens, but ultimately discarding the idea since rotating a map would make it less recognizable and unfamiliar to viewers.

\begin{figure}[t!]
	\centering
	\frame{\includegraphics[width=\columnwidth]{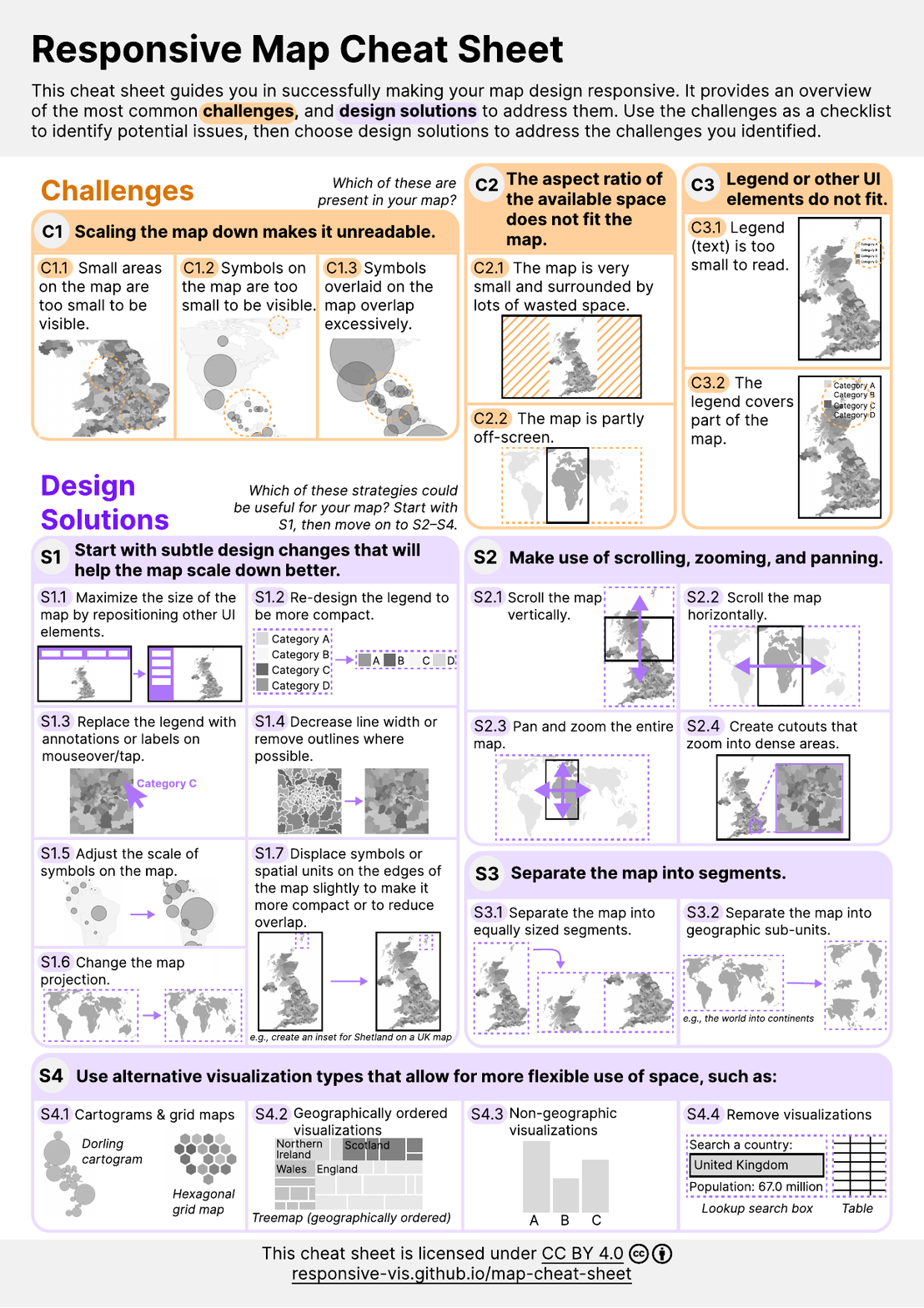}}
	\caption{Responsive Map Cheat Sheet, released under a CC BY 4.0 license and available in web page format or as a printable PDF at \sitelink.}
	\label{fig:cheatsheet}
\end{figure}

\section{Responsive Map Design Cheat Sheet}
\label{sec:cheatsheet}

We summarized the challenges and design solutions described in the previous section in an illustrated cheat sheet, displayed in \cref{fig:cheatsheet}.
Cheat sheets have been used in visualization before~\cite{wangCheatSheetsData2020} to make visualization techniques accessible to beginners.
Our aim is to make these strategies and design solutions available to anyone making responsive thematic maps, including visualization practitioners new to maps and/or responsive visualization, cartographers, and other professionals who make use of thematic maps such as urban planners.
The cheat sheet includes challenges (in orange) and design solutions (in purple).
Each challenge and solution is illustrated with a small schematic to make them easily understandable.
We designed the cheat sheet to reflect the overall hierarchy of strategies used by the experts we interviewed: design solutions are presented in four blocks, ranging from subtle changes (S1) to alternative visualizations (S4).

\begin{figure}[t!]
	\centering
    \includegraphics[width=\columnwidth]{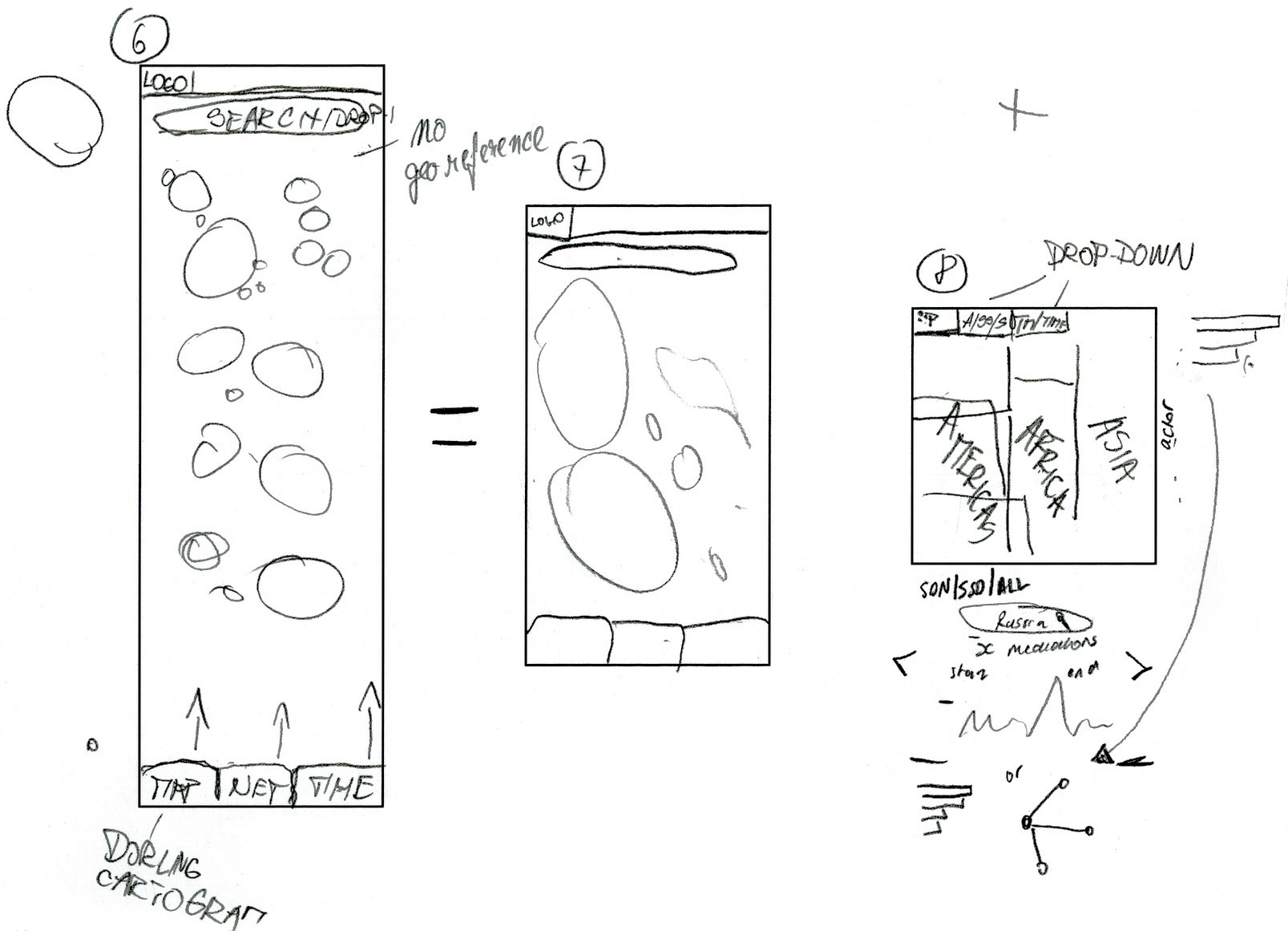}
	\caption{Sketches from our case study trialing the cheat sheet.
		Among other techniques, participants used cartograms (left) and segmented the map into continents (right).
	}
	\label{fig:workshop-sketch}
\end{figure}

We trialed this cheat sheet in a small case study with the project lead (no visualization background) and visualization developer of a map dashboard project.
Their project is part of a larger research project which has compiled a large global dataset that contains complex information on many countries, as well as links between countries.
They are now developing a map-based online dashboard focusing on certain aspects of this data.
At the time of the workshop, they had developed a prototype suitable for desktop screens.
We invited them to a 90-minute workshop, where we provided them with the cheat sheet as well as print-outs of differently sized rectangles, similar to the ones used on the Miro board in our workshops.
Over the course of the workshop, they successfully developed wireframe concepts for various sizes of their existing desktop-only map application.
They reported finding the cheat sheet helpful and initially used the list of challenges C1--C3 to identify issues in their map as it was scaled down.
They then used a variety of the proposed design solutions S1--S4 to address these issues, such as repositioning UI elements (S1.1), adjusting the scale of symbols (S1.4), segmenting the map into continents (S3.2), and using cartograms (S4.1).
\Cref{fig:workshop-sketch} shows some of their sketches for smaller screen sizes.

\section{Discussion}
\label{sec:discussion}

\textbf{Compared to previous studies} on responsive visualization design~\cite{kimDesignPatternsTrade2021,hoffswellTechniquesFlexibleResponsive2020}, our study shows strategies specific to challenges in responsive map design.
Interestingly, there seems to be a hierarchical approach, where designers typically follow a strategy that favors certain groups of design solutions over others (S1-S4).
The design solutions we compile are solutions to \textbf{map-specific challenges} that go beyond existing types of design patterns for other types of visualizations~\cite{kimDesignPatternsTrade2021}.
For example, while axes can be transposed for bar charts, rotating maps has been discouraged by participants in our study.
Likewise, the design pattern of \textit{changing encodings} leads to a whole set of geographic map visualizations as described in our S4.
However, the design rules, layout implementations, and implications for visualization literacy seem more complicated than simply a collection of techniques.
Consequently, our cheat sheet can be seen as a continuation of Kim et al's.
collection of design patterns~\cite[see Fig.~6]{kimDesignPatternsTrade2021}.

Our study also \textbf{confirmed earlier findings} on how designers and practitioners work~\cite{parsonsUnderstandingDataVisualization2022}, namely, the use of tacit and subjective knowledge and the inappropriateness of many of the prescriptive procedural schema.
Yet, despite the lack of a common \textit{design process}, our participants were aligned on many general strategies, used similar lines of reasoning, and not least---similar design solutions.
Our cheat sheet compiles these strategies and design solutions without being prescriptive about the \textit{process} designers follow.
While we concur that is true for expert designers, people new to visualization design might prefer more prescriptive models and guidelines.
We think our hierarchical strategies are able to cater to this need.
At least, these strategies highlight the complexity of decision making for responsive map design.
It is likely that we have barely scratched the surface of what is a common issue in visualization design: complex design decision making, constrained by devices, data, and users, with the help of guidelines.
Visualization designers must navigate this space and weight different factors.
Our strategies might help to bring more structure into this process and show that there are deliberate decisions in visualization design.
Moreover, our participants mostly agreed on these strategies which is evidence of our strategies' validity.
However, while we made an effort to select maps representative of the most commonly used thematic maps, our study only used two example maps with four participants each.
Studying additional map designs might reveal additional aspects of designers' decision-making processes, as well as additional design solutions.

We think \textbf{working with visualization designers} is an insightful, yet hugely underestimated, form of evaluation and research in visualization.
Especially as the number of professional visualization designers is growing and their work is more and more visible in the public and appreciated by analysts and communicators, we are able to obtain insights into application domains, their constraints, and behavior of a wide audience.
For example, mobile phone displays will continue to evolve, with foldable displays being one of the latest innovations; suddenly, visualizations need to adapt to square and mini displays.
But also, the fact that people (intuitively or deliberately) rotate their phones to rotate the content is unknown to conventional visualization design where we design for a fixed canvas.
Or, the fact that innovations in visualization for the public can be painstakingly slow out of fear of rejection---while at the same time having a huge potential for adaptation.

In order to support those new to responsive map design, a \textbf{guided design methodology or education activity}~\cite{huronIEEEVISWorkshop2020} could be developed based on our workshop methodology and the 12-rectangle sheet.
While participants in our pilot study reported a possible bias by seeing a list of possible geographic map visualizations, such a list or card deck could help ideation in an education context.
Further, we could create dedicated cheat sheets for different visualization design options and strategies.
However, our strategies are much more than a set of design patterns.
Many visualization cheat sheets describe charts and when to use them but cheat sheets can also indicate options and ``design pathways'', providing an overview of decisions (through decision trees) and thresholds.

Strategies and solutions can also \textbf{inform visualization tools}---for example, by providing challenge checklists, prescribing strategies, or providing suggestions based on the data, e.g., the map orientation.
Tools could also warn a designer if challenges manifest in the visualization, since issues in maps are not always obvious to less experienced map creators.
In terms of facilitating specific design solutions, tools could implement specific visualization designs, but they could also highlight existing implementations, including research prototypes that may be less well known.
For example, there is a wealth of schematized map designs (S4.1), many of which have been implemented as prototypes~\cite{hograferStateArtMap2020}.

This would help address a \textbf{major open challenge for practitioners}: how poorly supported many of these techniques are in current tools, libraries and software, despite the many improvements to tool support for generic responsive design in recent years.
For example, design solutions such as re-designing the legend (S1.2), displacing individual visual elements (S1.7), or segmenting maps into geographic sub-units (S3.2) will all generally require custom implementations.
Visualization developers who regularly build bespoke visualizations and interactions may be able to use these techniques by crafting implementations themselves, although time and budget constraints will limit this.
To visualization creators whose work is less custom, some design solutions will not be accessible at all without improved tool support.
This likely explains why many of our design solutions did not appear in prior work on responsive visualization that built on analyses of published visualizations---the more laborious to implement, the less likely a design will be found in the wild.

Finally, the majority of design solutions we identify have not been validated with \textbf{end users}, nor is there anecdotal evidence to draw from due to their lack of use in practice.
Conversely, there are validated techniques such as \textit{choriented maps}~\cite{gorteChorientedMapsVisualizing2022} that do not appear in our work.
User studies could validate the recommendations made by the designers in our study, resolve open questions where designers disagreed---e.g., to what extent pan and zoom solves responsiveness challenges in thematic maps---and also provide guidance on how to best apply these strategies.
For example, how well do users understand segmented maps (S3.1), how many segments are appropriate, and are there specific arrangements or map designs that work better than others?
What types of cartograms and grid maps (S4.1) are most easily understood by the general public?
Navigation techniques for maps have previously been investigated~\cite{burigatNavigationTechniquesSmallscreen2008}, but the study did not include techniques commonly used nowadays, so future studies might compare maps with pan and zoom (S2.3) to maps that can be scrolled in one direction only (S2.2 and S2.3).
Likewise, Brychtová~\cite{brychtovaExploringInfluenceColour2015} makes some recommendations on legend placement in choropleth maps; future user studies could build on this to design and compare compact legend designs (S1.2) as alternatives to common defaults.
Applying the design solutions proposed in this paper in real-world visualizations will undoubtedly raise many more questions on their application in various scenarios, design details, and implementation.

\section{Conclusion}

We introduced a set of seven challenges and 17 design solutions for responsive map design, based on eight workshops with experienced map designers and developers.
To make them accessible to visualization creators more widely, we compiled them into an illustrated cheat sheet detailing challenges, design solutions, as well as the overarching strategy for applying them.
However, our workshops also showed how complex, intuition-driven, and spontaneous decision-making in visualization design can be, limiting what can neatly be described as a formal process.
For the future, we hope that our work supports both visualization creators directly, and the development of new tools that make some of these strategies more accessible in practice.

\section*{Supplemental Materials}
\label{sec:supplemental_materials}
Our companion website at \sitelink\ provides web page and PDF versions of the cheat sheet, as well as full-size downloads of the Miro boards shown in Fig.~\ref{fig:miro}.





\bibliographystyle{abbrv-doi-hyperref-narrow}

\urlstyle{same}
\bibliography{cheatsheet_references}


\end{document}